\documentclass{emulateapj}

\usepackage{graphicx}
\usepackage{amsmath}
\usepackage{amssymb}
\usepackage{apjfonts}

\font\FermiSmallfont=cmssq8 scaled 1200

\def\LANLppthead#1#2{
\null 
\begin{center}\vskip -1.0truein{\hbox to 7.5truein {
\hfill
\vbox to 1in {\vfill \FermiSmallfont
              \hbox{#1}
              \hbox{#2}
              \vfill}
}}\vskip-0.0truein\end{center}}

\shorttitle{Mass Function of Dark Matter Halos}
\shortauthors{Warren, Abazajian, Holz \& Teodoro}

\begin{document}

\LANLppthead {LA-UR 05-4013}{astro-ph/0506395}
\title{Precision Determination of the Mass Function of Dark Matter Halos}

\author{Michael S.\ Warren\altaffilmark{1}, Kevork Abazajian\altaffilmark{1}, 
  Daniel E.\ Holz\altaffilmark{1} and Lu\'{\a i}s Teodoro\altaffilmark{1,2}}

\altaffiltext{1}{Theoretical Division, Los Alamos
  National Laboratory, Los Alamos, NM 87545}
\altaffiltext{2}{Department of Physics,
Kelvin Building,
University of Glasgow,
G12 8QQ,
Glasgow, 
Scotland, UK}


\begin{abstract} 
The predicted mass function of dark matter halos is essential in connecting
observed galaxy cluster counts and models of galaxy clustering to the
properties of the primordial density field.  We determine the mass function in the
concordance $\Lambda$CDM cosmology, as well as its uncertainty, using sixteen
$1024^3$-particle nested-volume dark-matter simulations, spanning a mass range
of over five orders of magnitude.  Using the nested volumes and single-halo
tests, we find and correct for a systematic error in the friends-of-friends
halo-finding algorithm.  We find a fitting form and full error covariance for
the mass function that successfully describes the simulations' mass
function and is well-behaved outside the simulations' resolutions.  Estimated
forecasts of uncertainty in cosmological parameters from future cluster count
surveys have negligible contribution from remaining statistical uncertainties
in the central cosmology multiplicity function.  There
exists a potentially non-negligible cosmological dependence (non-universality)
of the halo multiplicity function.
\end{abstract}


\section{Introduction}

Collapsed, virialized dark matter halos arise from density peaks in the
initially Gaussian primordial fluctuation field \citep{PS,BBKS}.  The abundance
of the most massive of these halos is exponentially sensitive to the amplitude
of the initial fluctuation field as well as the mean matter density, making
observed counts of their abundance extremely sensitive to these properties of
the density field, as well as the dark-energy dependent growth rate of the
density field (e.g., \citealt{haiman2001}).  Condensation of gas and 
formation of stars within these halos leads to the formation of galaxies
\citep{White1978}.  In addition, nonlinear clustering halo-models of dark matter and
galaxies require, as a basic component, the mass function (see
\citealt{cooray02} for a recent review).

The analytic theory of virialized object formation through collapse of
overdense regions as envisioned by \citet{PS} (hereafter PS) employs the fact
that a uniform overdensity in the Universe will evolve as a separate, closed
universe, initially expanding with the background, but then slowing and turning
around to collapse and virialize.  Since the abundance of overdensity peaks
only depends on the fluctuation scale $\sigma$, the abundance of halos can be
expected to be universal in these units.  Limitations of approximations in the
PS model, e.g., sphericity of collapse and spatial overlap, led to a
modification of the original form with parameters fit to simulations
(\citealt{sheth99}, hereafter ST).  Using the same simulations,
\citet{jenkins01} (hereafter J01) abandoned the form of the PS motivated mass
function to better fit the simulations' mass range, but their functional form
cannot be extrapolated beyond the range of the fit.  J01 found the mass
function in $\sigma$ to be approximately universal for several cosmologies at
the level of $\sim$15\%.  

Here, we present a quantification of the dark matter halo mass function and its
uncertainties with a suite of sixteen nested-volume $1024^3$ particle dark
matter simulations of the concordance $\Lambda$CDM cosmology.  We quantify the
uncertainty and full covariance of the mass function parameters.  Our
halo-finding methodology is given in \S\ref{fofcorrect}; mass function
determination and error analysis is presented in \S\ref{fit}, along with
implications for future cluster surveys' sensitivities;  we present our
conclusions in \S\ref{conclusions}.

\section{Numerical Simulations and Halo Mass Determination}
\label{fofcorrect}

We calculate the mass function of dark matter halos arising in a concordance
$\Lambda$CDM model by performing numerical simulations of structure growth and
halo formation.  We use the Hashed-Oct-Tree (HOT) algorithm, initially
described in \citet{warren93} and recently compared in detail with other
well-known cosmology codes in \citet{Heitmann04}.  We use a per-interaction
error bound based on the analysis in \citet{SalWar94a}. The fractional error
per interaction is set to be no worse than $10^{-5}$ at a redshift of 25,
increasing to $10^{-3}$ at redshifts of 5 and lower.  The number of timesteps
and Plummer smoothing lengths ranged from 1480 steps and $2.1 h^{-1}$kpc
(physical) for the highest resolution simulation, to 720 steps and $98
h^{-1}$kpc (comoving) for the largest volume.  Each simulation required about
$2\times10^{17}$ floating point operations, which can be computed in roughly 60
hours on a 1024 processor parallel computer using HOT.  Overall, the results
presented here required over four exaflops ($4\times10^{18}$ floating point
operations).

We model a universe with
flat geometry and parameters 
\begin{equation}
\mathbf p = (\Omega_M, \Omega_b,n,h,\sigma_8) = (0.3, 0.04, 1, 0.7, 0.9).
\label{cosmoparams}
\end{equation}
Initial conditions are derived from the transfer functions as calculated by
CMBFAST \citep{seljak96}.

\begin{figure}
\includegraphics[width=3.35in]{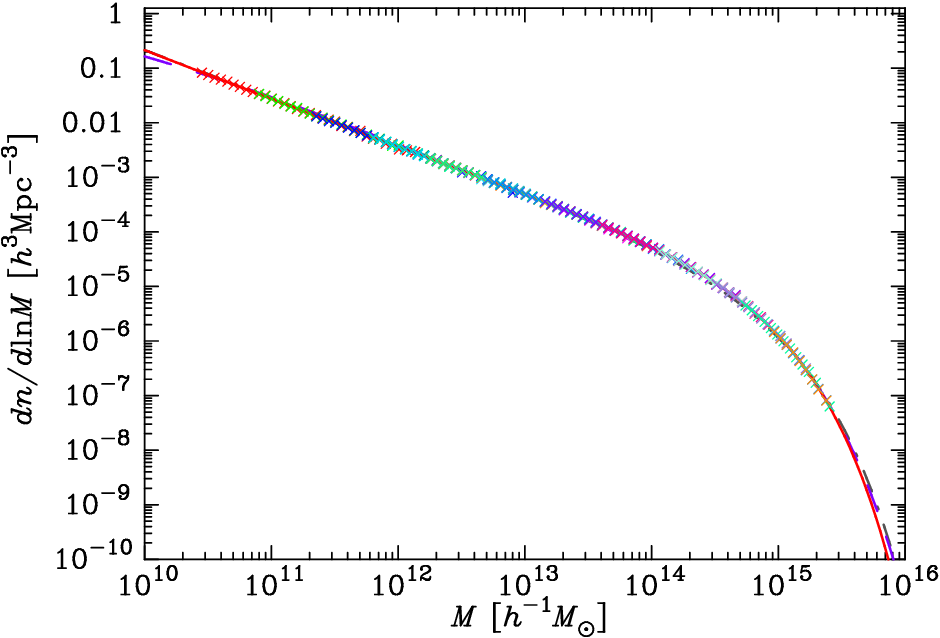}
\caption{ Shown are the central values of the binned mass functions from sixteen
  $1024^3$ simulations of the $\Lambda$CDM universe as crosses, with
  simulations in different colors. The best-fit form for the mass function we
  calculate is in solid (red), the Jenkins fit dashed (purple), and the
  Sheth-Tormen fit in dot-dashed (dark grey).  Goodness-of-fit is poorly judged
  on this extreme log scale; it is more clearly resolved in the linear residuals of Fig.~\ref{residuals}.
\label{logMF}}
\end{figure}

In order to simultaneously reduce Poisson error, verify consistency, and
resolve the widest mass-scale range available by our techniques, we employ
nested volumes with independent realizations of the chosen cosmology. We
simulated sixteen boxes of sizes 96, 135, 192, 272, 384, 543, 768, 1086, 1536,
2172, 2583 and 3072 $h^{-1}\rm Mpc$, with three realizations of 384 $h^{-1}\rm
Mpc$ size, and two each of 272 and 3072 $h^{-1}\rm Mpc$.  After the aggressive
requirement of a minimum of 400 particles per halo, we measure the mass
function over five orders of magnitude in mass scale; see Fig.~(\ref{logMF}).
The box size minimum is set by requiring the largest scale growth modes in the
volume to remain linear, while the the maximum is set by the halo particle
number requirement (i.e., very few massive halos in boxes of size $>
3\ h^{-1}\rm Gpc$ have more than 400 particles).  Our nested-volume approach
allows greater mass and length-scale resolution than pure particle number and
force resolution increase within a single simulation (e.g., \citealt{springel05}).

The friends-of-friends (FOF) method \citep{Frenk88} identifies a set
of particles which are spatially associated and contained within a
given isodensity surface defined by a linking parameter, $b$.
The linking length is defined as $h_{\rm link} = bn^{-1/3}$, where $n$ is the
number density of particles.

In practice, the number of particles per halo is not sufficient to
suppress statistical noise in the density field represented by the
finite number of particles.  This leads to a significant systematic error in the
estimated mass of a given halo, which depends on the number of
particles which represent the halo.  As an illustration of one aspect
of this problem, consider an isothermal density distribution of total
mass $M$ represented by $N$ particles within a sphere of radius R.
Each particle has mass $M_p = M/N$.  Choosing an FOF link length
$h_{\rm link}$ defines an isodensity surface with value
\begin{equation}
\rho(h_{\rm link}) = \frac{\alpha M_p}{(4\pi/3) h_{\rm link}^3},
\label{eq:isodens}
\end{equation}
where $\alpha$ is ``a constant of order 2'' \citep{Frenk88}.

Using the FOF method on a large number of sample isothermal halos with
parameters $M = R = 1$ and $h_{\rm link} = (N/1.25)^{-1/3}$ while varying $N$
leads to the results that are presented in Table~\ref{tab:mass-vs-n}.
It is clear from these results that there is a strong $N$-dependence in
the determined mass of the same underlying density distribution.
Another interesting result is that the mass values are converging to a
value which is significantly different from that implied from
Eq.~(\ref{eq:isodens}) with $\alpha=2.0$.  For large $N$, the correct
value for $\alpha$ is close to 3.1, which implies that the overdensity
of halos being found with the commonly favored link parameter of $b=0.2$
is about $280\rho_b$, rather than the often quoted value of $180\rho_b$
\citep{LaceyCole94, jenkins01, Whitehalo01, WhiteMF2002}.

\begin{table}
\begin{center}
\begin{tabular}{rllcrll}
\multicolumn{1}{c}{$N$} & \multicolumn{1}{c}{$M_{FOF}$} & \multicolumn{1}{c}{$\sigma$} & & \multicolumn{1}{c}{$N$} & \multicolumn{1}{c}{$M_{FOF}$} & \multicolumn{1}{c}{$\sigma$} \\
\hline
640000 & 0.3879 & 0.0014 & &  500 & 0.4704 & 0.0371 \\
 80000 & 0.3999 & 0.0037 & &  200 & 0.4964 & 0.0555 \\
 10000 & 0.4187 & 0.0092 & &  100 & 0.5222 & 0.0765 \\
  1250 & 0.4513 & 0.0244 & &   50 & 0.5586 & 0.1018 \\
\hline
\end{tabular}
\caption{The mean and variance of FOF mass vs $N$ for an ideal halo}.
\label{tab:mass-vs-n}
\end{center}
\vskip -.3cm
\end{table}

For a well-behaved halo finding method, one should be able to
sub-sample a simulation and derive the same mass function in the
regime where the statistics are robust.  In the case of FOF, this
means that one should be able to randomly pick one of every $n$
particles in the simulation, run the FOF method with a link length
$\sqrt[3]{n}$ times as large, and find the same distribution of halos.

However, due to the $N$-dependent statistical effects described above,
one obtains substantially different mass-functions from a sub-sampled
distribution.  While it is possible to use the binomial distribution
to determine an expression for the $N$-dependent corrections to the
FOF method in the case of an ideal isothermal density distribution,
halos found in simulations are complex objects which are not
well-described by such a smooth density distribution.  For this
reason, we have chosen to implement a correction which is calibrated
via the comparison of a simulation to the sub-sampled version of
itself.  Fortunately, the correction derived from such a procedure appears to
not depend strongly on the parameters of the simulation.  We find
a correction of the form
\begin{equation}
N_{corrected} = N(1 - N^{-0.6}),
\label{eq:FOFcorrection}
\end{equation}
for a halo with $N$ particles to do a reasonable job of correcting the
systematic error caused by the FOF method.

With this correction and the understanding of the behavior of FOF for
highly-resolved halos discussed above, for a link parameter $b=0.2$
(used for all further analysis in this paper) we find that the mean
density of halos with respect to the background in our largest volume
simulation is $260 \pm 50$, while in the smallest volume this value is
$340 \pm 60$.  For this reason, further analysis based on the results
described below should not assume that the overdensity of FOF halos is
constant with respect to scale.  The process of defining a halo
remains the largest uncertainty in defining the mass function.  New
algorithms for halo finding based on an overdensity criterion rather than
the use of FOF will be necessary for straightforward comparisons with
observational data.

\begin{figure*}
\centerline{\includegraphics[height=7.25in,angle=270]{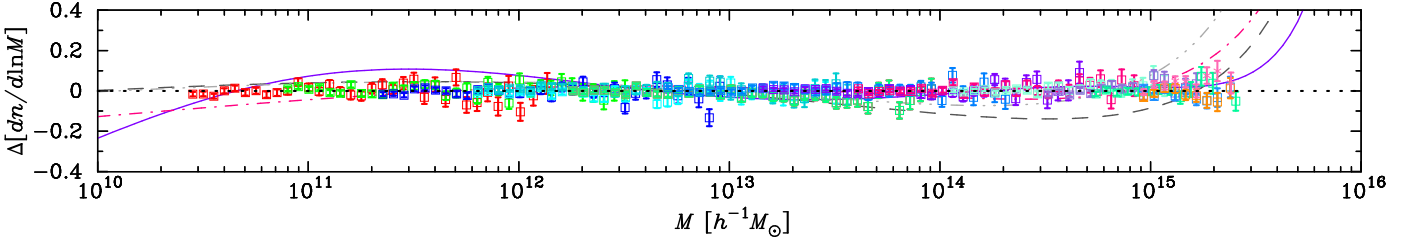}}
\caption{ Shown are the residuals from the binned simulation data to the fit presented in
  this work as square data points of different colors per simulation.  The
  Jenkins fit is the solid (purple) line, ST original fit the dashed (dark
  gray) line, the ST fit with parameters $A,a,p$ free with dot-dashed line
  (red), and the ST fit with $a,p$ free and amplitude $A$ set to require all
  dark matter in halos as a triple-dot-dashed line (light gray).   The binned
  mass function from the Virgo Hubble Volume simulation are the asterisk points
  with errors (pink). 
\label{residuals}}
\end{figure*}

\section{The Mass Function}
\label{fit}

The relation between the multiplicity function $f(\sigma)$ and mass function
$n$ is
\begin{equation}
f(\sigma) = \frac{M}{\bar\rho} \frac{dn}{d\ln \sigma^{-1}},
\label{multiplicity}
\end{equation}
where $\bar\rho$ is the mean matter density of the universe.  The general shape
of the mass function is well described by the PS and ST forms for the
multiplicity function.  We therefore use a similar form for the mass function
as that of ST, with a disentanglement of the power-law small-mass regime from
the exponential cut-off, and removal of the arbitrary collapse scale
$\delta_c$, which was nevertheless modified by the ellipsoidal collapse scale
in the ST model.  The minimal-parameter multiplicity function required to fit
the mass-function shape measured in the simulations is
\begin{equation}
f(\sigma) = A \left(\sigma^{-a} + b \right) e^{-c/\sigma^2},
\label{lanlfit}
\end{equation}
with parameters ${\mathbf q} = (A,a,b,c)$.  This form is related to the
multiplicity function resulting from the barrier shape ansatz of
\citet{st2002}, but without an associated change of the exponential.

Using, Eq.~(\ref{lanlfit}), we calculate the extended maximum
likelihood, $\lambda({\mathbf q})$ for the Poisson data of the mass function
found in the combined set of simulations~\citep{rpp},
\begin{equation}
\ln\lambda({\mathbf  q}) = -\sum_{j=1}^{N_{\rm s}}\sum_{i=1}^{N_j}\left[\mu_{ij}({\mathbf q}) -
  n_{ij} + n_{ij}\ln\frac{n_{ij}}{\mu_{ij}({\mathbf q})}\right],
\label{likelihood}
\end{equation}
of $N_{\rm s}$ simulations containing $N_j$ bins, where $n_{ij}$ is the number
of halos in mass bin $i$ of simulation $j$. In bins where $n_{ij}=0$ the last
term is zero.  The predicted number of halos in bin $i$ of simulation $j$
between masses $m_{ij1}$ and $m_{ij2}$ is
\begin{equation}
\mu_{ij}({\mathbf q}) = V_j \int_{m_{ij1}}^{m_{ij2}}{ \frac{dn({\mathbf
      q})}{dM}\ {d M}},
\label{expected}
\end{equation}
where $V_j$ is the volume of the simulation.  The extended maximum likelihood
is appropriate for Poisson data sets such as the mass function, and allows for
an estimation of the goodness-of-fit for a given model, with the number of
degrees of freedom for the multinomial nature of the distribution given by
$N-k-1$, where $N$ is the total number of bins and $k$ is the number of fitted
parameters.

The mass function from all simulations is binned with minimum logarithmic width
of $0.05$, with the last bin expanded to include a minimum of
400 halos.  Higher mass halos have negligible statistical weight and are
discarded.  The low-mass end requires 400 particles per halo, and the FOF
determined masses are corrected as described in \S\ref{fofcorrect}.

Sample variance can be a consideration when attempting to use the most massive
halos available in a small volume (e.g., see \citealt{Hu:2002we}).  The sample
variance of halos of mass $M$ in a cubic volume of side length $L$ such as our
simulations is $b^2(M)\sigma^2(L)$, where $b(M)$ is the halo bias and
$\sigma(L)$ is the fluctuation amplitude in the cubic volume.  For our
simulation volume samples, the maximum sample variance is $0.2\%$ of the
Poisson error, which occurs for the most massive halo bin in our smallest
volume simulation.  Sample variance is therefore negligible here.

We perform an analysis of the likelihood, Eq.~(\ref{likelihood}), for the
multiplicity function, Eq.~(\ref{lanlfit}), to obtain
the simulations' results for the parameters $\mathbf q$ free.  The best-fit of
the multiplicity function is found with parameters
\begin{eqnarray}
A &=& 0.7234\pm 0.0043\ {\rm  (stat.)} \pm 0.003\ {\rm (sys.)},\cr
a &=& 1.625\phantom{0}\pm 0.019\phantom{0}\ {\rm  (stat.)} \pm 0.009\ {\rm (sys.)}, \cr
b &=& 0.2538\pm 0.0031\ {\rm  (stat.)} \pm 0.002\ {\rm (sys.)}, \cr
c &=& 1.1982\pm 0.0055\ {\rm  (stat.)} \pm 0.002\ {\rm (sys.)},
\label{bestfitpars}
\end{eqnarray}
with corresponding 1$\sigma$ statistical errors, and estimated systematic
uncertainty.  The statistical error covariance is
\begin{equation}
{\mathbf C} =
\begin{pmatrix}
1.82  & 5.80  & 0.0730 & 2.04\phantom{0} \\
\hdotsfor{1}  &  \!36.7\phantom{6}   &  4.35\phantom{00} & 9.85 \phantom{0}\\
\hdotsfor{2}  &  0.958\phantom{0}  & 0.854 \\
\hdotsfor{3} &  2.97\phantom{0}\\
\end{pmatrix}\cdot 10^{-5}.
\end{equation}
Error analysis was performed with the MINUIT package of CERNLIB.  The $\chi^2$
per degree-of-freedom (DOF) for the best fit is $\chi^2/{\rm DOF} = 524.5/429$,
indicating a larger scatter than expected from the Poisson statistics.  

The uncertainty and resulting scatter in the sampling of a given overdensity by
the FOF halo finder likely contributes to the higher than expected $\chi^2/{\rm
  DOF}$.  To assess this source of systematic uncertainty, we vary the FOF
correction Eq.~(\ref{eq:FOFcorrection}) within ranges fit by single simulations, as
well as alternate binnings of the mass function, which sample this distribution
differently.  The estimated systematic uncertainties are subdominant to
statistical, though not negligible.

The residuals of the binned mass functions from all simulations to the expected
number of halos in the given bin [Eq.~(\ref{expected})] of mass-function best
fit are shown in Fig.~\ref{residuals}.  In this figure, we also plot the binned
mass function from the Virgo Hubble Volume simulation~\citep{jenkins01}, which
we derive from their particle data identically to our simulations
(although the initial power spectrum for those data are not precisely
the same as our simulations).  

For comparison, we also fit the ST form for the multiplicity function.
With amplitude free of the constraint that all dark matter be within
halos, the ST parameters are $A=0.3405\pm 0.0004$, $a=0.7183\pm 0.001$
and $p= 0.157 \pm 0.003$, with $\chi^2/{\rm DOF} = 1987/430$.  From
the derived $\chi^2/{\rm DOF}$ and Fig.~\ref{residuals}, modifications
of ST form and the J01 form clearly provide a poor fit to the
$\Lambda$CDM mass function.  However, the ST parameters are similar to
that fit to the bias from simulations of \citet{seljak04} using the ST
form as given in \citet{Mandelbaum:2004mq}.  Therefore, the
peak-background split ansatz of the relation between halo biasing and
the mass function may be consistent \citep{mo96,sheth99,sheth01}.
Although more can be done to verify this statistical consistency, it
is beyond the scope of this {\it Letter}.

\begin{figure}
\includegraphics[width=3.35in]{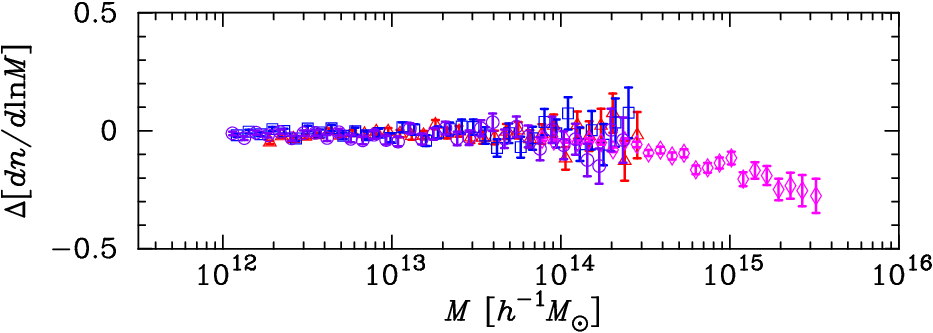}
\caption{ Residuals are shown for simulations with varied cosmological parameters
  with that cosmologies' mass function using the multiplicity function predicted
  from the central cosmology.  Cosmological parameter are held constant, except
  triangles are for $\sigma_8 = 0.8$; squares are for $\Omega_m =0.2$; circles
  for $h=0.6$ and $\sigma_8 = 0.8$; diamonds are for $\Omega_m =0.2$ and
  $\sigma_8 = 1.2$.
\label{alt_residuals}}
\end{figure}

To test the claim of the multiplicity function's universality in consistently
describing varying cosmologies, we find the residuals for the mass function
derived from the $\Lambda$CDM simulations' multiplicity function versus
simulations done for varying cosmological parameters within $\Lambda$CDM.  Four
separate $1024^3$ simulations were run with parameters as those in
Eq.~(\ref{cosmoparams}) except, a $384\ h^{-1}\rm Mpc$ box with $\sigma_8 =
0.8$; a $384\ h^{-1}\rm Mpc$ box with $\Omega_m =0.2$; a $329\ h^{-1}\rm Mpc$
box with $h=0.6$ and $\sigma_8 = 0.8$; and a $1536\ h^{-1}\rm Mpc$ box with
$\Omega_m =0.2$ and $\sigma_8 = 1.2$.  Residuals are shown in
Fig.~\ref{alt_residuals}.  The models are consistent within the $5\%$ level,
except for the $\Omega_m =0.2$ and $\sigma_8 = 1.2$ model, which shows
departures at $20\%$.  The departure may be more than a Poisson fluctuation,
revealing a statistically significant dependence of the multiplicity function
on cosmology.  Quantifying this potential dependence is warranted, but beyond
the scope of this work.

A potential implication of this work is the calculation of the sensitivities
of current and future cluster surveys to properties of the primordial
perturbations and the dark-energy-dependent growth of these perturbations.
Some of the original work on quantifying parameter sensitivities of future
cluster surveys considered the bias associated with mass function uncertainties
\citep{Holder2001,haiman2001}, as well as in more recent work
\citep{Battye:2003bm}.  However, the uncertainties in the predicted dark matter
halo mass function are often entirely ignored \citep{Wang:2004pk,Wang:2005vr}.

To test the potential implications of dark matter halo mass function
uncertainties on forecasts for parameter sensitivity, we use the Fisher matrix
approach of \citet{Holder2001} for sensitivities of the South Pole Telescope
(SPT) Sunyaev-Zeldovich Effect cluster survey, with redshifts determined by the
Dark Energy Survey (DES), and a minimum mass cutoff of $M_{\rm min} =3 \times
10^{14}\ h^{-1} M_\odot$.  Sky coverage is $4000\ \rm deg^2$ and cluster counts
are binned in redshifts $\Delta z=0.05$ from $z=0$ to $z=2$.  We have also
calculated and included the Fisher information from forecasts for the Planck
mission parameter sensitivities from the cosmic microwave background (CMB)
$TT$, $TE$ and $EE$ correlations~\citep{Eisenstein:1998hr}.  In this test, we
have not included scatter in the mass-temperature
relation~\citep{Levine:2002uq}, nor uncertainty in this
distribution~\citep{Lima:2005tt}, but have also not included mass-observable
self-calibration~\citep{Majumdar:2002hd}, which can be compensating effects.
We find that the inferred uncertainties forecast for parameters best
constrained by an SPT+DES cluster survey beyond the CMB constraints, namely the
equation of state, $\Omega_m$ and neutrino mass, have forecast errors increased
by, at most, approximately 1\% in the error.  Therefore, the mass function
presented here has its uncertainties sufficiently well determined so that they
are negligible compared to observational uncertainties.

\section{Conclusions}
\label{conclusions}

We have measured the shape and quantified the uncertainty in the
predicted theoretical mass function of dark matter halos in the
$\Lambda$CDM cosmology with sixteen separate $1024^3$ dark matter
structure formation simulations.  We found and corrected a systematic
error in halo mass determination by the friends-of-friends halo
finder.  The canonical \citet{sheth99} and \citet{jenkins01} forms of
the mass function are inconsistent with the $\Lambda$CDM mass function
at $\sim$10\% level at intermediate masses, and $>$30\% at the highest
masses.

Combining the work presented here with uncertainties in halo biasing will be
necessary for improving work on quantified applications of the halo model
\citep{vandenbosch03,Abazajian:2004tn}.  The peak-background split
model for the relation between halo bias and the mass function appears
consistent, although this should be studied in further detail.

Analyses of current and future cluster surveys' sensitivities to the primordial
perturbation spectrum and dark energy are not greatly dependent on inherent
statistical uncertainties in the predicted dark matter halo mass function.
The use of high-redshift cluster count surveys will require quantification of
the predicted evolution of the mass function, as well as any cosmological
dependence of the halo mass multiplicity function.  Simulation efforts may be
tailored for a specific survey's requirements in this regard.  In concert with
observations, this will help elucidate the nature of dark energy, neutrino
mass, cosmological structure and galaxy formation.

\acknowledgments

We would like to thank Josh Frieman, Salman Habib, Katrin Heitmann, Alexia
Schulz, Uro\v s Seljak, Ravi Sheth, Jochen Weller and Simon White for useful
conversations.  We thank the Virgo Supercomputing Consortium for making
available to us their Hubble Volume data.  This work was performed under the
auspices of the U.S. Dept. of Energy, and supported by its contract
\#W-7405-ENG-36 to Los Alamos National Laboratory.  The computational resources
were provided by the LANL open supercomputing initiative and the Space
Simulator Beowulf cluster.


\bibliography{lss}

\begin{thebibliography}{33}
\expandafter\ifx\csname natexlab\endcsname\relax\def\natexlab#1{#1}\fi

\bibitem[{Abazajian {et~al.}(2005)}]{Abazajian:2004tn}
Abazajian, K., {et~al.} 2005, \apj, 625, 613

\bibitem[{{Bardeen} {et~al.}(1986){Bardeen}, {Bond}, {Kaiser}, \&
  {Szalay}}]{BBKS}
{Bardeen}, J.~M., {Bond}, J.~R., {Kaiser}, N., \& {Szalay}, A.~S. 1986, \apj,
  304, 15

\bibitem[{Battye \& Weller(2003)}]{Battye:2003bm}
Battye, R.~A., \& Weller, J. 2003, Phys. Rev., D68, 083506

\bibitem[{{Cole} \& {Lacey}(1994)}]{LaceyCole94}
{Cole}, S., \& {Lacey}, C.~G. 1994, \mnras, 271, 676

\bibitem[{Cooray \& Sheth(2002)}]{cooray02}
Cooray, A., \& Sheth, R. 2002, Phys. Rept., 372, 1

\bibitem[{Eidelman {et~al.}(2004)}]{rpp}
Eidelman, S., {et~al.} 2004, Phys.\ Lett.\ B, 592, 1

\bibitem[{Eisenstein {et~al.}(1998)Eisenstein, Hu, \&
  Tegmark}]{Eisenstein:1998hr}
Eisenstein, D.~J., Hu, W., \& Tegmark, M. 1998, Astrophys. J., 518, 2

\bibitem[{Frenk {et~al.}(1988)Frenk, White, Davis, \& Efstathiou}]{Frenk88}
Frenk, C.~S., White, S. D.~M., Davis, M., \& Efstathiou, G. 1988, \apj, 327,
  507

\bibitem[{{Haiman} {et~al.}(2001){Haiman}, {Mohr}, \& {Holder}}]{haiman2001}
{Haiman}, Z., {Mohr}, J.~J., \& {Holder}, G.~P. 2001, \apj, 553, 545

\bibitem[{Heitmann {et~al.}(2004)Heitmann, Ricker, Warren, \&
  Habib}]{Heitmann04}
Heitmann, K., Ricker, P., Warren, M.~S., \& Habib, S. 2004, \apjs, in press,
  astro-ph/0411795

\bibitem[{{Holder} {et~al.}(2001){Holder}, {Haiman}, \& {Mohr}}]{Holder2001}
{Holder}, G., {Haiman}, Z., \& {Mohr}, J.~J. 2001, \apjl, 560, L111

\bibitem[{Hu \& Kravtsov(2003)}]{Hu:2002we}
Hu, W., \& Kravtsov, A.~V. 2003, \apj, 584, 702

\bibitem[{Jenkins {et~al.}(2001)}]{jenkins01}
Jenkins, A., {et~al.} 2001, \mnras, 321, 372

\bibitem[{Levine {et~al.}(2002)Levine, Schulz, \& White}]{Levine:2002uq}
Levine, E.~S., Schulz, A.~E., \& White, M.~J. 2002, Astrophys. J., 577, 569

\bibitem[{Lima \& Hu(2005)}]{Lima:2005tt}
Lima, M., \& Hu, W. 2005, Phys. Rev. D., submitted, astro-ph/0503363

\bibitem[{Majumdar \& Mohr(2003)}]{Majumdar:2002hd}
Majumdar, S., \& Mohr, J.~J. 2003, Astrophys. J., 585, 603

\bibitem[{Mandelbaum {et~al.}(2004)Mandelbaum, Tasitsiomi, Seljak, Kravtsov, \&
  Wechsler}]{Mandelbaum:2004mq}
Mandelbaum, R., Tasitsiomi, A., Seljak, U., Kravtsov, A.~V., \& Wechsler, R.~H.
  2004, \mnras, submitted, astro-ph/0410711

\bibitem[{{Mo} \& {White}(1996)}]{mo96}
{Mo}, H.~J., \& {White}, S.~D.~M. 1996, \mnras, 282, 347

\bibitem[{{Press} \& {Schechter}(1974)}]{PS}
{Press}, W.~H., \& {Schechter}, P. 1974, \apj, 187, 425

\bibitem[{Salmon \& Warren(1994)}]{SalWar94a}
Salmon, J.~K., \& Warren, M.~S. 1994, J. Comp. Phys., 111, 136

\bibitem[{{Seljak} \& {Warren}(2004)}]{seljak04}
{Seljak}, U., \& {Warren}, M.~S. 2004, \mnras, 516

\bibitem[{{Seljak} \& {Zaldarriaga}(1996)}]{seljak96}
{Seljak}, U., \& {Zaldarriaga}, M. 1996, \apj, 469, 437

\bibitem[{{Sheth} {et~al.}(2001){Sheth}, {Mo}, \& {Tormen}}]{sheth01}
{Sheth}, R.~K., {Mo}, H.~J., \& {Tormen}, G. 2001, \mnras, 323, 1

\bibitem[{{Sheth} \& {Tormen}(1999)}]{sheth99}
{Sheth}, R.~K., \& {Tormen}, G. 1999, \mnras, 308, 119

\bibitem[{{Sheth} \& {Tormen}(2002)}]{st2002}
---. 2002, \mnras, 329, 61

\bibitem[{Springel {et~al.}(2005)}]{springel05}
Springel, V., {et~al.} 2005, \nat, 435, 629

\bibitem[{van~den Bosch {et~al.}(2003)van~den Bosch, Mo, \&
  Yang}]{vandenbosch03}
van~den Bosch, F.~C., Mo, H.~J., \& Yang, X. 2003, \mnras, 345, 923

\bibitem[{Wang {et~al.}(2005)Wang, Haiman, Hu, Khoury, \& May}]{Wang:2005vr}
Wang, S., Haiman, Z., Hu, W., Khoury, J., \& May, M. 2005, Phys. Rev. Lett., in
  press, astro-ph/0505390

\bibitem[{Wang {et~al.}(2004)Wang, Khoury, Haiman, \& May}]{Wang:2004pk}
Wang, S., Khoury, J., Haiman, Z., \& May, M. 2004, Phys. Rev., D70, 123008

\bibitem[{Warren \& Salmon(1993)}]{warren93}
Warren, M.~S., \& Salmon, J.~K. 1993, in Supercomputing 1993 (Los Alamitos:
  IEEE Comp.\ Sci.), 12

\bibitem[{{White}(2001)}]{Whitehalo01}
{White}, M. 2001, \aap, 367, 27

\bibitem[{White(2002)}]{WhiteMF2002}
White, M. 2002, \apjs, 143, 241

\bibitem[{{White} \& {Rees}(1978)}]{White1978}
{White}, S.~D.~M., \& {Rees}, M.~J. 1978, \mnras, 183, 341

\end{thebibliography}
\bibliographystyle{apj}

\end{document}